\documentclass[manuscript=review,american,utf8]{beilstein}
\usepackage{xspace}
\usepackage{caption}
\usepackage{physics}
\usepackage{xcolor}
\usepackage{etoolbox}
\usepackage{hyperref}
\DeclareCaptionLabelFormat{adja-page}{\hrulefill\\#1 #2 \emph{(previous page)}}
\usepackage{subfig}

\newcommand{\rezA}{\AA$^{-1}$}

\newcommand{\rezAs}{\AA$^{-1}$ \,}

\newcommand{\mums}{$\mu m$\,}
\newcommand{\mus}{$\mu s$}
\newcommand{\muss}{$\mu s$\,}
\newcommand{\Jm}{J/m$^2$}
\newcommand{\Jms}{J/m$^2$\,}

\begin{document}
\renewcommand{\baselinestretch}{1.1} 

\title{Laser fragmentation in liquid – constructing a generic reaction map}

\author*{Anton Plech}{anton.plech@kit.edu}
\affiliation[1]{Institute for Photon Science and Synchrotron Radiation, Karlsruhe Institute of Technology (KIT), Postfach 3640, D-76021 Karlsruhe, Germany}

\author[1]{Yogesh Pokhrel}

\author{Meike Tack}
\affiliation[2]{Technical Chemistry I and Center for Nanointegration Duisburg-Essen (CENIDE), University of Duisburg-Essen, 45141 Essen, Germany}

\author{Matteo Levantino}
\affiliation[3]{ESRF}

\author[2]{Sven Reichenberger}

\maketitle

\begin{abstract}
Excitation of absorbing colloids in liquid by pulsed laser irradiation creates highly non-equilibrium states of matter that relax via defined pathways of thermal and non-thermal dissipation channels. The structural fingerprint of these channels can be probed by ultrafast {\it in situ} x-ray scattering methods and classified as a function of temporal delay of excitation excitation density, or specifically laser fluence. We study the example of photo-excitation of a gold colloid by picosecond laser pulses at the interband absorption band at 400 nm in water. By quantifying lattice temperature, crystalline fraction, particle sizes and the structural response of the water around the nanoparticles a reaction map is constructed that allows to pinpoint heating, particle melting, water bubble formation and finally particle fragmentation to form nanoclusters of predominant sizes of <  3 nm for applications in theranostics, photonics or catalysis. 
\end{abstract}

\keywords{laser processing in liquids, laser fragmentation in liquid, time-resolved probing, x-ray scattering, nanoparticle melting, nanobubble formation}


\section{Introduction}

In laser synthesis and processing of colloids (LSPC) the rapid deposition of energy in a strongly confined volume of material leads to non-equilibrium states of matter to trigger structural and chemical transitions that lead to nanostructures that are desirable in applications. In particular, processing of colloids in liquids has emerged as a conceptually simple approach to produce nanoparticles of defined size, shape, or composition \cite{zhang17}. Current goals for tailoring nanostructures are creating ultrasmall clusters with improved catalytic or photonic performance \cite{ziefuss21}, achieving non-equilibrium alloyed particles, or taking advantage of precise size and shape distributions \cite{xu06,ament12,gonzalez17} for diagnostic or therapeutic applications \cite{wang10,guo17}. In a first step laser-based synthesis is often achieved by pulsed-laser ablation of solid targets immersed in liquids \cite{fojtik93,Shih2018}. As a second step, the further laser-processing of the obtained colloidal nanoparticles can add a better definition of size and/or morphology  by fragmenting the produced colloid (LFL) with large or multimodal size distribution \cite{mafune02,plech05cpl,besner06,ziefuss18} or by fusing nanoparticles to larger, size-defined spheres after a melting (LML) process \cite{wang10,swiatkowska17}. While the application of pulsed lasers on liquid-submerged surfaces or in a colloid (LFL/LML) is conceptually simple, mastering of the process to tailor specific product properties (e.g. particle size, crystallinity and crystal structure, composition, oxidation state or morphology) requires a functional understanding of the underlying dynamics and mechanisms. Central for understanding processes such as LFL or LML is the description of light absorption, energy transduction pathways, and thermal kinetics. 

\begin{figure*}[ht] 
\caption{Sketch of possible pathways for nanoparticle excitation and fragmentation by ultrashort laser pulses starting with particles suspended in a liquid like water at ambient temperature T$_A$. The excitation close to the melting temperature T$_m$ leads to lattice heating, partial or full melting, and eventual bubble formation during cooling. Depending on particle temperature, resp. fluence, partial fragmentation, repulsion by the inverse Leidenfrost phenomenon or full fragmentation for temperature > T$_m$ can take place. Nonlinear effects, such as strain-induced fragmentation or near-field effects can take place.}
\label{mech}
\includegraphics[width=15cm,keepaspectratio]{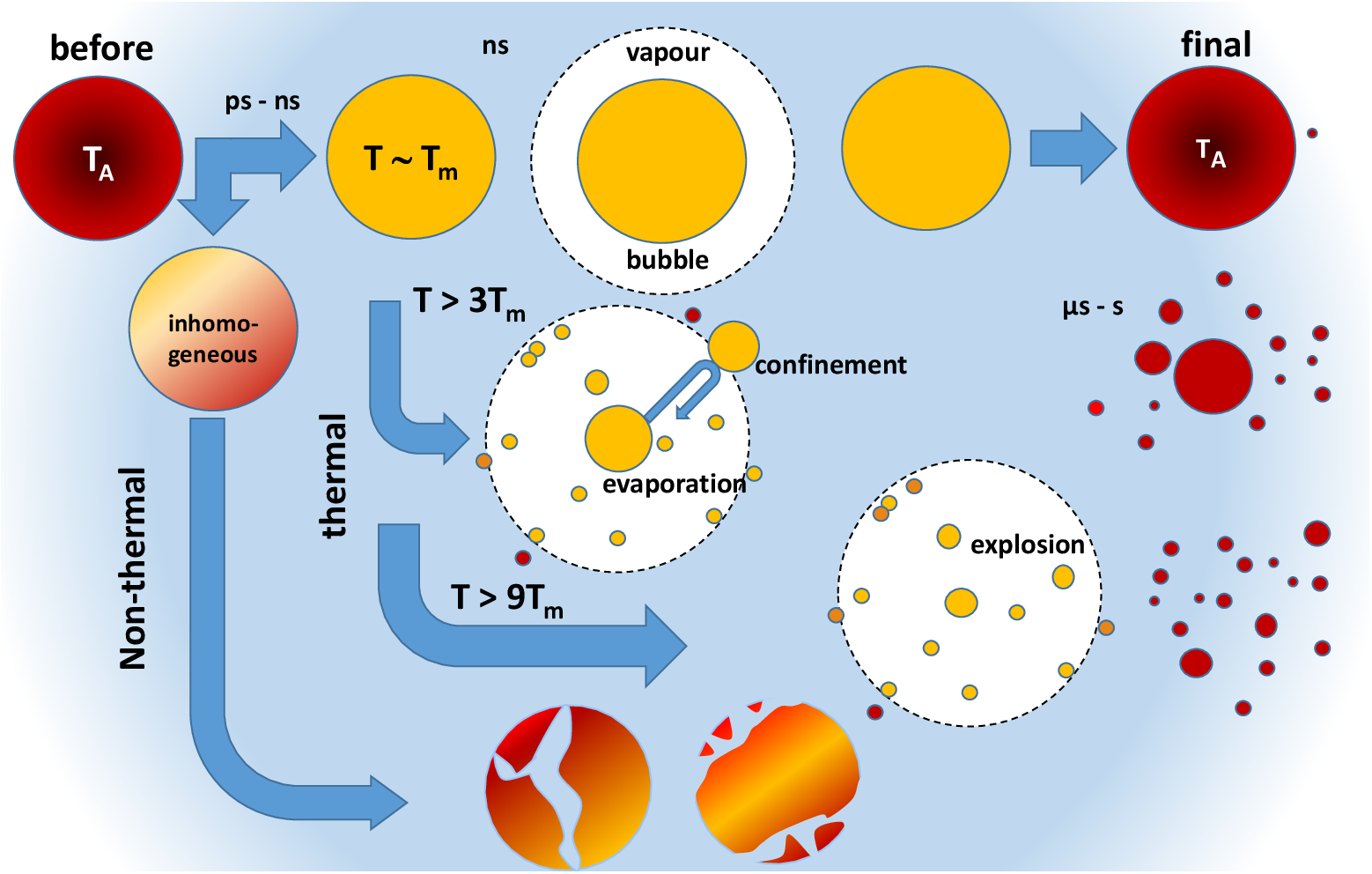}
\end{figure*}

A way to rationalize the elements of structure formation is to assign an internal energy to each subsystem, which can finally be parametrized by a resulting transient temperature function of the respective subsystem. Each of these subsystems (be it free electron temperature, lattice temperature, liquid temperature, or charge separation) is described by interacting with the others either in a linear way (such as by electron-phonon coupling or heat conduction) or by surpassing thresholds that lead to non-linear response (such as forces by near-field effects, lattice melting, evaporation, or structural reshaping). This allows to classify the mechanisms that lead to structure formation as well as correlate final products to the energy density and dynamics. A basic classification as a scheme of temporal and spatial dynamics is given in fig. 1 a for pulsed laser irradiation of colloids in liquid. On the level of the single isolated  particle, irradiation at ambient temperature leads to lattice heating \cite{hartland03soft,plech04prb}, in most cases via electron excitation and lattice coupling \cite{link00rev,wilson02,guzelturk20}. With sufficient provision of laser energy as sketched in fig. \ref{mech} the particles may undergo melting, release the heat into the liquid \cite{wilson02} and finally return back into the non-excited state almost unchanged. One central aspect is that for excitation that is faster than any structural response and faster than dissipation from the particles, the energy uptake can be quantified by linear properties of the material, like absorption cross section for photons, heat capacity, and kinetics of heat flow. In that case, the laser fluence (as energy per unit area) can serve as a useful parameter to classify the thermal effects. In such a thermal reaction, increasing the absorbed energy to a multiple of the melting temperature (fig. \ref{mech}) can lead to thermal decomposition, as described by evaporation or  phase explosion \cite{huang21,plech24nano}. The produced small clusters undergo further interaction and may eventually be stabilized in size. Apart from that, in non-thermal mechanisms the fluence (or alternatively the lattice temperature) will no longer be sufficient to describe structure formation. Rather charge separation or development of strain can aid in the destruction of the particles, with shapes of the fragments being determined by the peculiarity of energy distribution or energy gradients in space and time, as shown by the lower pathway in fig. \ref{mech}. This has been discussed in terms of charge-induced fragmentation of particles that can undergo Coulomb instability if the surface charge surpasses the cohesion of a liquid particle or by field-induced particle disruption \cite{yamada06,werner11}, which can be facilitated by plasmonic nanoparticles \cite{plech06np,delfour15,voss19,bongiovanni21}. On the other hand, strain confinement in nano- or microparticles can also exert inhomogeneous forces on nanoparticles that can lead to fragmentation without the necessary thermal energy to overcome evaporation and phase explosion barriers \cite{ihm19,spellauge23,pokhrel26}.

Since all these previously described fragmentation mechanisms stem from the same physical correlations and parametrization, they can be unified in a generic reaction map. However, such a map has not yet been developed since tracing the exact fragmentation mechanism on ps- to ns- timescales for the given energy uptake (as a function of applied laser fluence, absorption cross-section of the particles, etc.) is far from trivial. Within this report, we describe a pump-probe experiment on a gold colloid that uses x-ray scattering over a broad scattering range to characterize the structural relaxation as function of fluence and temporal delay after excitation. The experiments follow a true {\it in situ}, single-shot data acquisition scheme to map the pathways of energy relaxation and discuss the mutual interaction of these pathways.  Although the absolute time scales for structural relaxation of the particles and the surrounding liquid will vary upon changing metals or liquid or by size and shape of the colloid, the relations of the thermal kinetics, liquid response, and particle fragmentation are generic up to a certain point. The classification in a reaction map as function of time delay and laser fluence should be a good starting point to find similarities across different metal colloids and understand thermal processes like thermal fragmentation to be contrasted to other mechanisms.

\section{Experimental methods and samples}

The experiment is designed as a single-shot experiment, where every ensemble of particles is exposed to one single pulse of picosecond radiation at a defined laser fluence and delay with a resolution of 60 ps. The fluence is used as fundamental parameter instead of other possible quantification of laser intensity, such as power density resp. irradiance, because we believe that in the present case the pulse length is too long for nonlinear effects, but yet short enough to deliver the energy into the particles before relaxations set in, as will be discussed later. 

\subsection{Colloid synthesis and characterization}

\begin{figure*}[ht] 
\caption{Size distribution and electron microscopy images of selected colloids samples produced by seeded chemical synthesis (CTAB) and LAL followed by optional size separation. The lines of size distribution correspond to data from disc centrifugation, while the bar graphs include histograms from TEM analysis. }
\label{samples}
\includegraphics[width=16cm,keepaspectratio]{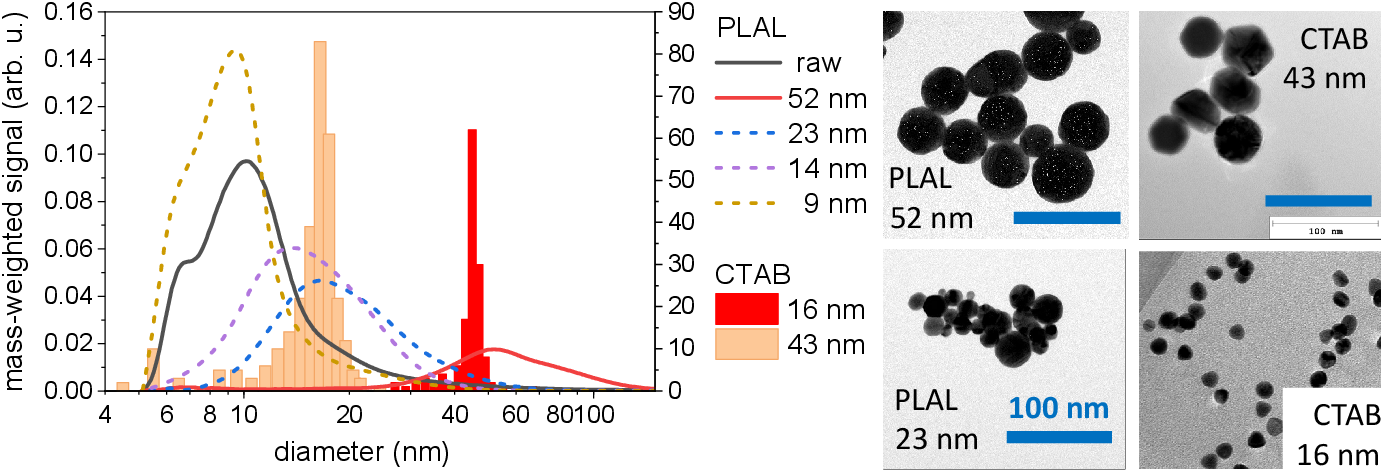}
\end{figure*}

The colloidal gold nanoparticles in the range from 7 to 167 nm has been synthesized by a set of different methods that each produce nanoparticles of a defined size range but result in different properties, such as crystallinity, size distribution width, or ligands. (i) Pulsed laser ablation in liquid has been used to produce a native ("raw") suspension of gold particles that have been fractionated in size in a second step by repeated centrifugation of the pellet and supernatant independently, as described in \cite{ziefuss20,plech24nano}. In brief, a Nd:YAG laser (Ekspla, 10 ps, 1064 nm, 9.6 mJ per pulse, 100 kHz) has been focused into a batch chamber onto a gold target for 10 min. The colloid has been stabilized by a NaOH/NaCl solution. These samples have a marked size dispersion after fractionation of 13-17 \% root mean square deviation (rms) and are polycrystalline \cite{Shih2018,reich19nanoscale}, but ligand-free except for the Stern layer formation (ii) Seeded synthesis has been used to prepare single-crystalline particles of excellent size dispersion (5-8 \% rms), but slightly faceted shapes that resemble cuboids. In brief, seed particles are prepared by rapidly reducing gold hydrochlorate (HAuCl$_4$) in a solution of cetyl trimethyl ammonium bromide (CTAB) by NaBH$_4$.  In a second step growth of larger particles around the seeds is initiated by stepwise addition to a HAuCl$_4$ solution in CTAB and mild reduction by sodium ascorbate according to established protocols \cite{Zheng14}. A repetition of the seeding protocol with the diluted colloid from the second step allowed even larger particles to grow. Platinum particles of 7 nm have been grown using the same recipe by using PtCl$_4$.  (iii) Citrate reduction of HAuCl$_4$ following the Turkevich method \cite{turkevich51}, but with variation of the gold concentration and the gold-to-citrate ratio \cite{kimling06} produced spherical colloid with higher size dispersion of 12 - 15 \% rms. The particles are polycrystalline.
Sample analysis is done by small-angle x-ray scattering (SAXS) at the SAXS instrument of the KIT Light Source \cite{simmler22}, by transmission electron microscopy, by a disc centrifuge and/or by UV/Vis optical spectroscopy. Exemplified sample distributions are shown in fig. \ref{samples}.

The colloid in each case was used at a concentration between 60-100 mg/l for CTAB, 150 -200 mg/l for LAL particles and 200 mg/l for citrate particles. A free liquid jet pointing downward was realized using a syringe pump and an opened x-ray capillary (Hilgenberg) as an orifice. The jet speed 0.72 m/sec. allowed one to replace the liquid in the interaction area between each laser shot at 1 kHz.

\subsection{Synchrotron pump-probe x-ray scattering}

The pump-probe experiment used a dedicated setup at the European Synchrotron Source ESRF, beamline ID9 \cite{levantino21}, where a synchronized laser (Coherent Ti:Sa regenerative amplifier; frequency doubled to 400 nm; pulse length 1 ps) is focused onto the jet from the side at an angle of 8 degrees to about 0.23 mm full width at half maximum (fwhm). The center of the excited path is probed by x-ray pulses from the synchrotron perpendicularly to the jet with a spot size of 0.04 mm. 

Pairs of synchronized pulses were prepared by a set of x-ray choppers and synchronizing the laser oscillator to the RF frequency of the synchrotron. Delays with 5 picosecond steps up to 1 ms were realized by combining the phase-lock synchronization to a defined pulse picking of the amplifier stage. The x-ray energy was chosen to be at 15 keV at the maximum of the single-line emission of the undulator that delivers a band width of 1 \% without using an additional monochromator. Scattered x-ray radiation is detected on a two-dimensional detector (Rayonix 170HS) with 84 \mums spatial resolution. The covered solid-angle range 2$\Theta$ is varied using a detector distance of 640 mm and 220 mm to cover a range of the scattering vector q$= 4 \pi / \lambda \cdot \sin(2\Theta/2)$ between 0.008 and 6 \rezA. Original data is available from the ESRF repository \cite{repo1,repo2}.

At each setting of delay and fluence, a pair of two images is taken that each acquires scattering from 2000 to 20000 shots with the delay first set to a negative value (- 2ns) and then to the desired value. Thus, the negative delay can serve as reference measurement of x-ray scattering before the laser pulse reaches the sample, and drifts, fluctuations, and background scattering are largely suppressed.  
The laser fluence is defined by carefully measuring the focal spot size \cite{plech22ACS} and changing the intensity by a combination of rotatable waveplate (1/2 $\lambda$ retardation) and a Glan-Laser polarizer. 

\subsection{Data reduction and analysis}

\begin{figure} 
\caption{Composite difference scattering intensities $\Delta$I at delay $\tau = 1 $\muss after laser excitation as compared to the static scattering before laser arrival ($\tau$<0). The experimental results (scatter plots) are simulated by a linear combination of structural contributions as discussed in the text. Data is shifted by a constant vertical offset for clarity.}
\label{sq}
\includegraphics[width=8cm,keepaspectratio]{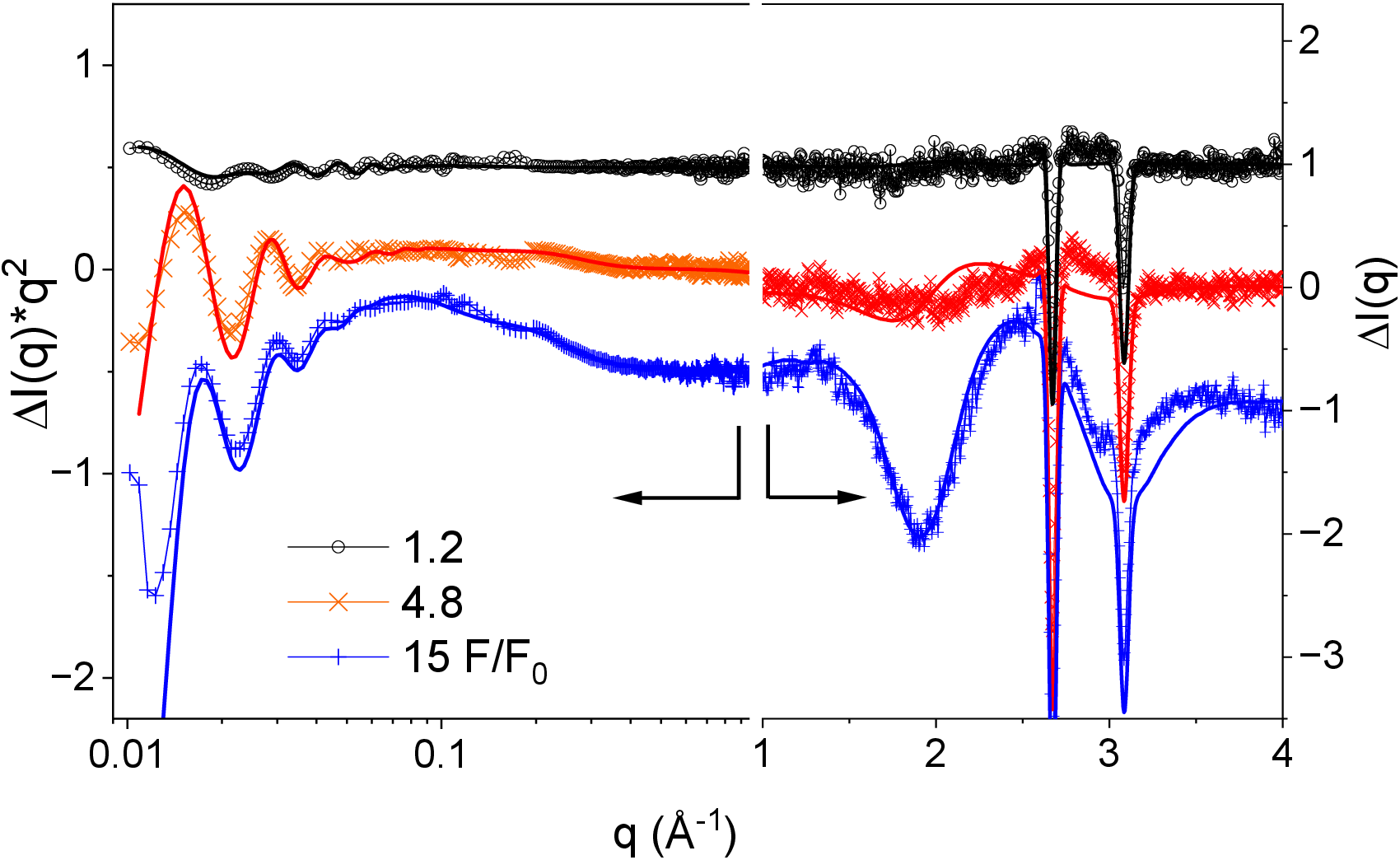}
\end{figure}

The pair of images from the x-ray detector are corrected for dark counts, space-angle coverage, and polarization anisotropy according to established protocols \cite{plech07jsr} and invalid pixels are masked. The reference image at negative delay is subtracted from the data image by normalizing both at a q value around 4 \rezA (WAXS), which corresponds to an isosbestic point of scattering change of water and the nanoparticles, respectively around 0.5 \rezAs for SAXS. An azimuthal integration produces a difference function as shown in fig. \ref{sq} with positive and negative contributions induced by laser excitation.

The SAXS region up to 0.5 \rezAs is modeled by a linear combination of scattering functions that include the initial spherical particle (which is subtracted upon fragmentation), a new spherical particle as a result of size change and a positive contribution from newly formed fragments. The latter are modeled by the generalized function \cite{sztucki07}, by including two hierarchy levels of sizes, while the former is modeled by the analytical sphere form factor \cite{plech24nano}. Bubble formation can also be included by adding form factors of core-shell particles with a shell whose density can be reduced from water to vacuum \cite{kotaidis05apl}. The details are described in the supporting information of ref. \cite{plech24nano}. 

The WAXS region is analyzed at the position of the powder rings (Bragg peaks of the gold fcc lattice) at 2.67 \rezAs and 3.14 \rezAs in terms of peak position, width and integrated intensity. The (111) peak is used to determine a potential lattice expansion $\Delta a$ (unit cell size a)  that reflects the incremental thermal strain (da) as given by the thermal expansion coefficient $\alpha$ and the temperature rise dT:

\begin{eqnarray}
\frac{1}{a} da & = & \alpha(T) dT \nonumber \\
dQ(T) \quad & = & \quad \frac{4 \pi}{3} \cdot R^3 \cdot \rho \/ c_p(T) \cdot dT \quad = \quad \sigma_{abs} \cdot dF 
\end{eqnarray} \label{eqheat}

with the temperature rise being caused by the input of thermal energy dQ. As both $\alpha$ and $c_p$ are temperature-dependent, the set of differential equations needs to be solved numerically to predict the fluence-dependent lattice expansion. It is useful to know that the temperature dependence of $\alpha(T)$  and $c_p(T)$ nearly cancels out, which leads to a well linear relation between the observed lattice expansion $\Delta$a/a and the applied fluence F. The absorption cross section $\sigma_{abs}$ may become be fluence dependent, if optical properties change within the duration of the irradiating laser pulse. One such non-linear effect is plasmon bleaching, which is important close to the SPR around 532 nm, while at 400 nm no losses of absorbed energy have been reported \cite{wang20ch,plech22ACS}. To calibrate the fluence scale despite possible ambiguities in the absorption cross section we introduced the threshold fluence F$_0$ at which the extrapolated lattice expansion reaches the value of the maximum lattice expansion at melting of up to 1.8 \% for bulk gold. The melting point depression reduces the melting temperature from 1336 K of bulk gold moderately to about 1290 K with a particle diameter of 10 nm \cite{buffat76} or 1050 K for 5 nm \cite{wang24}. The maximum expansion is then adapted to the size-dependent melting point, for example, calculated to be 1.71 \% for 20 nm particles. 

The scattering from water is stronger than the gold powder rings by 2 orders of magnitude, as the gold concentration in the colloid is as small as 10$^{-5}$ in volume. Thus, even weak changes in the water structure due to energy transduction from gold can be detected in scattering changes that are stationary in q to the first approximation. Changing the thermodynamic parameters of volume, pressure and temperature then leads to two linearly independent functions of the scattering change, as described earlier \cite{cammarata06,kjaer13,hanson26}. Consequently, using tabulated functions for $\partial S / \partial T$ and $\partial S / \partial V$ as given in the following expression:

\begin{equation} \label{dsrho2}
\Delta S(\tau, q) = \pdv{S}{T} \biggl|_{\rho} \cdot \Delta T + \pdv{S}{\rho}\biggl|_{T} \cdot \Delta \rho
\end{equation}

the time-dependent changes in the state of water can be quantified by fitting the amplitudes $\Delta T$ and $\Delta V$ to the scattering change over a wide range in q. 

The modeling of all scattering changes of the colloid as a function of fluence and delay is exemplified in fig. \ref{sq}. At q below 0.1 \rezAs the signal is dominated by oscillations of the SAXS form factor of the particles together with a reduction of intensity at < 0.02 \rezA , which indicates particle destruction. At the position of the Bragg reflections sharp negative changes are seen, which indicate the loss of long-range order through melting and fragmentation. A broad change of scattering is visible with minima close to 2 and 3 \rezAs and a maximum close to 2.5 \rezA. This modulation reflects the change of the liquid water structure from heating and volume change. 

The difference curves are analyzed semi-automatically by a Python script that integrates the 2D data, subtracts background in a first step at positions other than the Bragg peaks to allow for Gaussian fitting of the peaks. In a second step, the derived peak positions are used to carefully fit the difference curves by a combination of contributions from nascent small fragmented particles by the generalized SAXS description, the liquid scattering and a change of powder peak positions and intensity relative to the values of the data at negative delay from the first step. The change of the initial particle size is derived by manual fits of the entire SAXS/WAXS curve to the full balance of particles removed and appearing as described in \cite{plech24nano}. A consequence of automated fits is that outliers with a modified parasitic  background are not removed in any case.   




\section{Results and discussion}

The excitation and relaxation steps of metal colloids involve several mechanisms, which have been documented exemplarily earlier, such as interband or plasmonic absorption \cite{link00rev}, electron-phonon coupling \cite{hartland04pccp,arbouet03}, lattice disordering by the thermal Debye-Waller factor \cite{guzelturk20} and by thermally induced melting \cite{plech04prb,plech24nano} as well as a number of non-linear channels opening up with  laser excitation shorter than electron-phonon coupling \cite{plech06np,werner11,voss19}.

\begin{figure}
\caption{Lattice dynamics of 47 nm CTAB gold particles as function of laser fluence and delay with 
a) showing the lattice expansion and (111) peak intensity as function of applied fluence together with a calculation of the predicted expansion with adapted threshold for melting F$_0$. 
b) Lattice expansion as function of delay for selected fluences together with a model of exponential cooling decay (lines) and 
c) (111) peak intensity as function of delay together with a model of intensity recovery.
d) (111) Bragg peak intensity of 52, resp. 60 nm particles as function of lattice expansion at delay 60 ps for different gold colloids as prepared by laser ablation (PLAL) and ligand-stabilized chemical reduction (CTAB). The dashed vertical line marks the maximum lattice expansion of gold, when the melting point is reached. The solid red line shows a prediction of intensity reduction of a bulk crystal due to thermal vibrations (Debye-Waller effect). 
e) Compilation of fluences F$_0$ as function of particle diameter and particle synthesis method. The lines are prediction of the caloric lattice expansion with fluence by using the calculated absorption cross section from Mie theory (dashed line) and a correction by assuming a finite time resolution of the experiment of 60 ps and considering partial cooling of the lattice during the duration of the probe pulse (full red line). }
\label{expans}
\includegraphics[width=16cm,keepaspectratio]{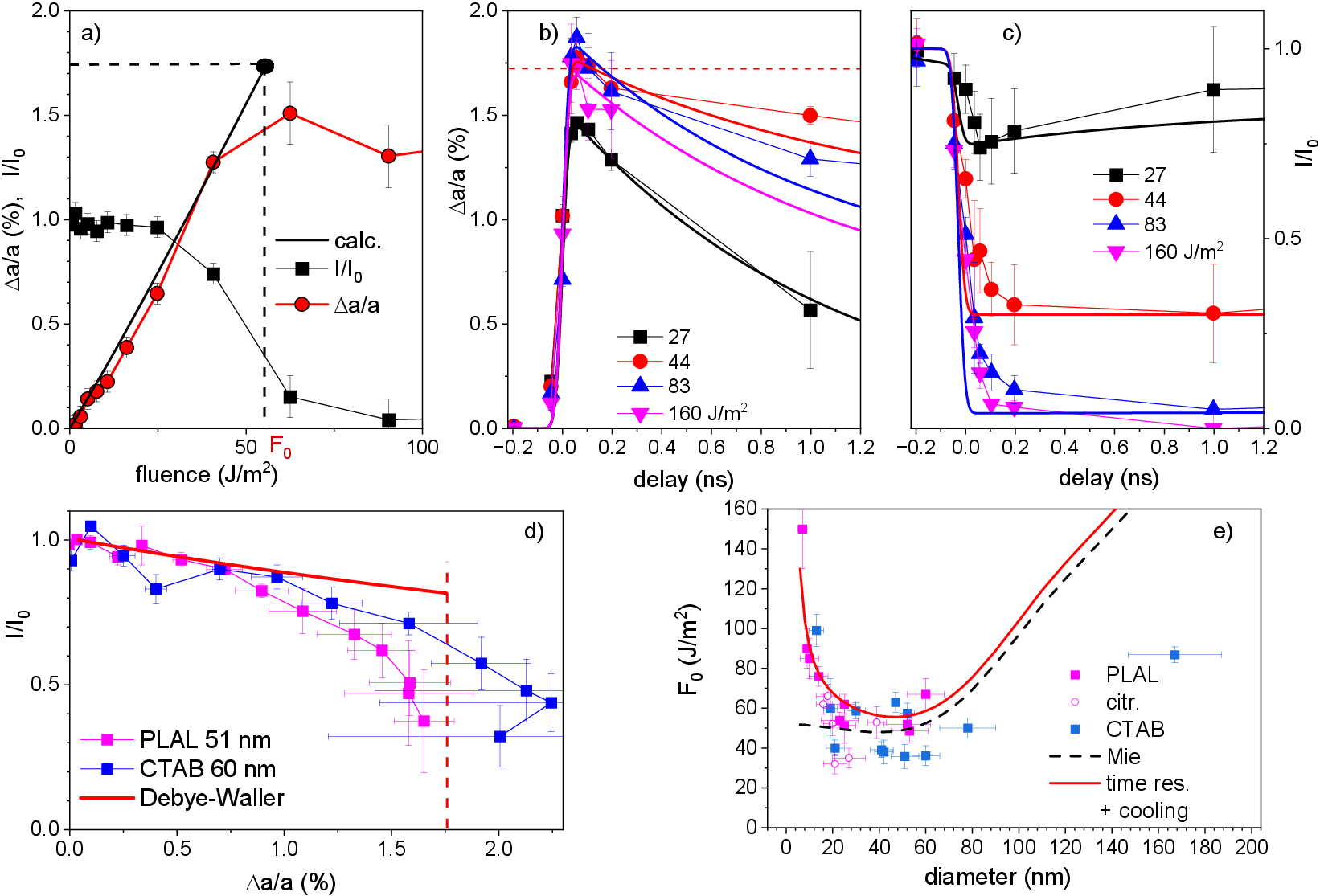}
\end{figure}

\clearpage

\subsection{Lattice dynamics, heating and melting}

Within this discussion, we establish the energy scale of the processes by quantifying thermal expansion and reduction of lattice ordering as a function of applied laser fluence. A gradual increase of laser fluence leads to an increase in lattice expansion at a delay of 60 ps, as seen in fig. \ref{expans} a). The delay of 60 ps is chosen as the shortest delay within the available time resolution of the experiment. At the same time, it ensures that the absorbed laser energy has been fully transformed into lattice thermal energy, given an electron-phonon coupling time of 2-10 ps \cite{arbouet03,guzelturk20,plech24njp}. Coherent lattice vibrations may be present, but with periods in the range of 10 - 20 ps for nanoparticles of 30 - 60 nm diameter \cite{hartland03soft,arbouet06,kotaidis12} coherent strain due to these vibrations will be averaged over the time resolution. Consequently, we assume that the lattice expansion is proportional to the temperature increase far away from the melting point. Fig. \ref{expans} b) shows that as a result of cooling of the nanoparticles the expansion decays exponentially with a lifetime of 1.2 ns for CTAB particles of 47 nm diameter. The expansion develops within the experimental time resolution.

On the other hand, when temperatures close to the melting temperature of gold are reached this assumption is expected to not hold any more, as melting will reduce the diffraction intensity of the powder peaks relative to the still crystalline volume \cite{warren}. Thus, at higher fluence the peak intensity will drop upon melting together with a lower-than linear increase of expansion with fluence, as only the still solid (and colder) particles contribute. This is seen in fig. \ref{expans} a) where at about 40 \Jms the peak intensity drops quickly beyond the pure thermal disorder contribution (Debye-Waller factor at the melting point is about 0.8). The observed lattice expansion of the remaining solid particles therefore remains in a range below 1.74 \%, in agreement with the size-dependent at the melting point. The linear range, where no intensity losses are seen, can then be used to calibrate the fluence F$_0$ required to reach the melting point by using the integrated form of eq. 1 with scaled absorption efficiency (see below). Here, the intersection between the predicted lattice expansion and the measured data with the maximum expansion of gold results in F$_0 =$ (54$\pm$5) \Jm. 

It is interesting to note that at fluences close to the melting fluence (44 and 83 \Jms) the intensity of the Bragg (111) peak of the gold lattice (fig. \ref{expans} c) does not drop with the same speed as the lattice expansion (fig. \ref{expans} b). In fact, at 83 \Jms the lowest peak intensity is only reached at a delay of 1 ns while the most dominant changes in I/I$_0$ already occurred within the first 200 ps . We interpret this as a melting transition that is partly delayed with respect to the lattice temperature increase. It has previously been discussed that melting as a first-order phase transition involves the formation of nuclei, preferably at heterogeneous nucleation sites \cite{mo18,staechelin21} with moderate superheating. Only at the superheating limit homogeneous melting speeds up the process to melt the lattice within a few picoseconds \cite{rethfeld01}. Melting times can be longer than 100 ps \cite{arefev22} with a recent study in gold films showing that the ultimate limit for superheating is 14 times the static melting point temperature if superheating is achieved in a few picoseconds \cite{white25}.   A consequence of the superheating is that the lattice expansion may increase for temperatures above the melting temperature, and thus overshoot the predicted maximum expansion. In fig. \ref{expans} d) the reduction of peak intensity is shown as a function of lattice expansion. The expected reduction of peak intensity as function of temperature and thus of observed lattice expansion is shown as the solid red line following equation \ref{dwf} as computed by the relation

\begin{equation} \label{dwf}
I/I_0 = exp(-1/3 \langle u^2\rangle \cdot q^2)
\end{equation}

with the mean square displacement <u$^2$> being a function of temperature. The parameterization is given in ref. \cite{gao99}. Clearly, for CTAB particles the expansion proceeds beyond 1.75 \% (predicted for 51 nm particles) but not for PLAL particles. The reason may be the crystalline morphology. While CTAB particles as mostly single crystalline particles (crystal sizes matches the particles size), PLAL particles are typically polycrystalline with many defect planes that result from rapid quenching after synthesis \cite{Shih2018,reich19nanoscale}. Therefore, heterogeneous melting is kinetically favored for PLAL particles, while CTAB particles can be overheated beyond the melting temperature. 

We note that the calibrant fluence F$_0$ is derived directly from the experiment in order to correct for accuracy limits of the experiment, such as non-Gaussian laser beam shape or the caustic of the laser beam inside the liquid jet. Currently, a comprehensive understanding of fluence-dependent absorption efficiency is not available, with earlier results showing bleaching at plasmon resonance \cite{magnozzi19,plech22ACS}  or even slight enhancement at interband wavelengths for gold \cite{wang20ch}. Here we compile the results from several experiments conducted at different experimental campaigns with the same technique for various nanoparticle batches. Fig. \ref{expans} e) shows the extracted values of F$_0$ as a function of the particle size and the synthesis method together with a prediction using static absorption cross sections. In static light interaction, the absorption cross section per particle volume is fairly constant at the interband absorption at 400 nm \cite{siems11} and can be used for the quantification of nanoparticle concentration \cite{molina26}. This would predict a constant F$_0$ independent of the particle size, which is visible as a dashed line. For nanoparticles with a diameter below 70 nm F$_0$ is hence expected to be around 50 \Jms. However, with the present temporal resolution limit and the fact that for small particles the lattice cooling towards the liquid environment could already set in within tens of picoseconds, the measured lattice expansion underestimates the maximal reached expansion. Therefore, we have convolved the prediction with the time resolution and the calculated cooling kinetics \cite{plech04prb,plech07jsr} to find a much higher (apparent) F$_0$ for particles smaller than 20 nm (see red line marked as "time res. + cooling" in Fig. 4e). Given the error margins of the experiment, the experimentally derived values of $F_0$ match well the prediction of Mie theory and lifetimes. Note that there is a spread in fluence across different sample batches, which may be related to differences in size or size distribution. In particular the PLAL particles show a wider size distribution, while CTAB particles are characterized by a different degree of faceting from cuboid shapes to nearly spheres \cite{plech22ACS,plech24nano}.  Overall, the expected melting point as a function of fluence can be well predicted and shows a minimum around particle sizes of about 60 nm as previously discussed \cite{katayama14}.
 
\subsection{Cooling and bubble formation}

\begin{figure*}[ht] 
\caption{Amplitudes of the water scattering change $\Delta \rho$ scaling with the bubble volume and $\Delta T$ as function of delay for a) CTAB gold particles of 17 nm and 52 nm diameter at intermediate fluences of 800 and 1600 \Jm , respectively and b)  $\Delta \rho$ for a range of particles sizes including 7 nm Pt nanoparticles at a fluence of 2700 \Jms. $\Delta \rho$ for 17 nm particles is compared to the (111) peak intensity in B and shown as function of normalized fluence F/F$_0$ at several delays in d). The error bars are derived from the variation of repeated measurements.}
\label{bubble}
\includegraphics[width=17cm,keepaspectratio]{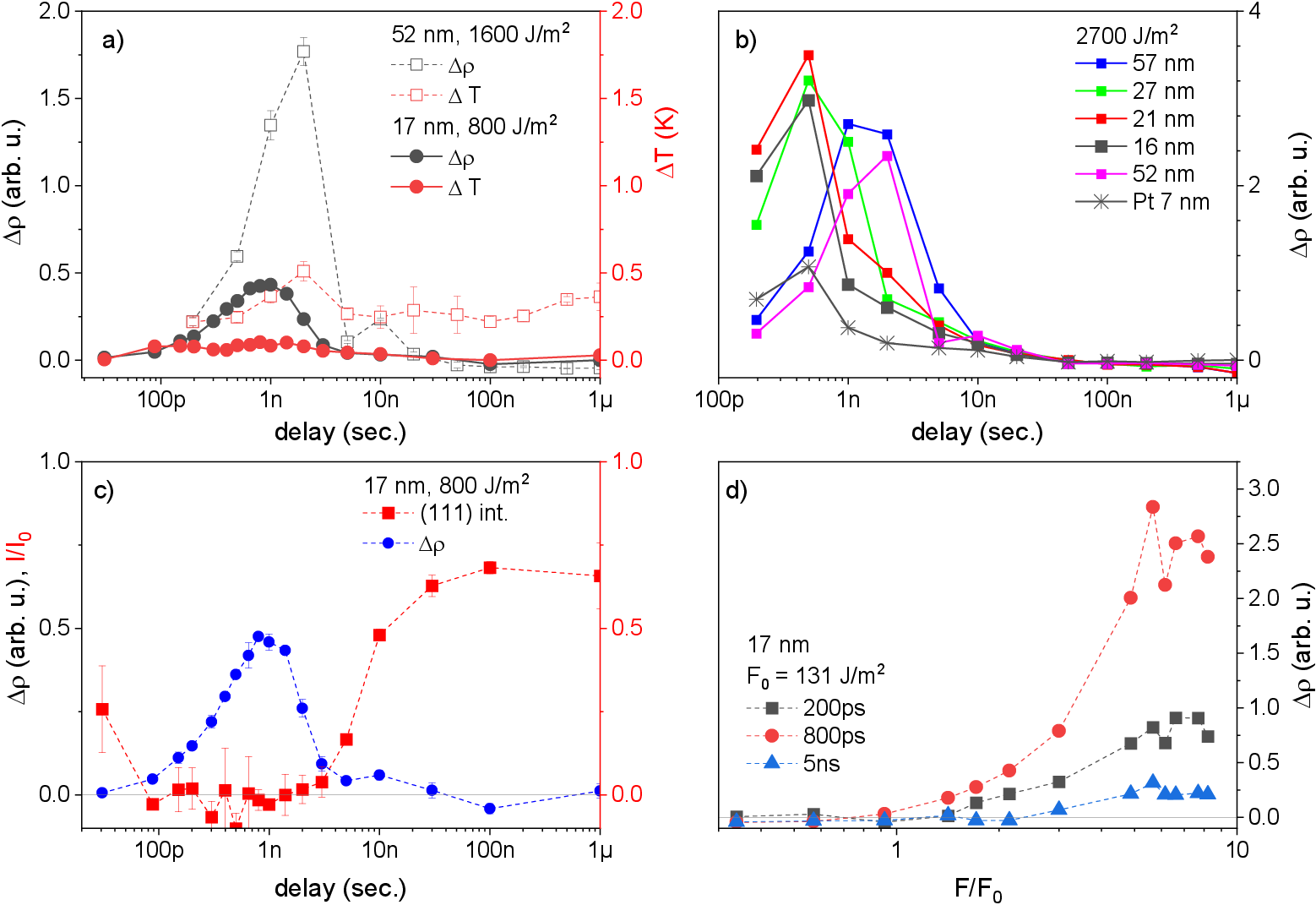}
\end{figure*}

The temperature of the particles, parameterized by lattice expansion, decays after the initial heating with a time constant of 1.1 ns at 27 \Jms to a double exponential decay of 1.1 and 30 ns at higher fluences, as derived from the data in fig. \ref{expans} b). This cooling is explained by heat transfer across the gold-water interface, where the thermal conductance of water and the finite conductance G across the interface compete in determining the time scale. Defining a critical interface conductance $G_c$ \cite{wilson02} :

\begin{equation}
G_c = \frac{3 C_f K_f}{C_p} \frac{1}{R}
\end{equation}

with $C_f$ and $C_p$ being the volumetric heat capacity of the fluid and the particle respectively, and $K_f$ the thermal conductivity of the fluid gives an estimate of whether G or the thermal conduction in water would dominate. The dependence on the particle radius with 1/R shows that the interface conductance dominates for the smallest particles, with typical conductances have been found to be between 5 and 200 MW/m$^2$/K \cite{wilson02,plech04prb,plech22ACS,arefev25}. Arefev et al. \cite{arefev25} show by detailed molecular dynamics simulations that the interface conductance of uncapped gold particles in water starts at 110  MW/m$^2$/K and decreases with temperature followed by a strong drop to 5-40 MW/m$^2$/K where the water in the immediate vicinity of the gold surface reaches the critical point. At high enough energy deposited in the particles the rise of the water temperature close to the critical point, resp. crossing the spinodal line leads to homogeneous nucleation of vaporization, where the time constant for formation of a vapor bubble does not depend on nucleation probability but only on bubble hydrodynamics (Rayleigh Plesset equation, RPE). Consequently, bubble formation has been  previously observed\cite{kotaidis05apl,kotaidis06jcp,siems11}. Bubble dynamics can be roughly described by the competition of vapor pressure against water pressure and water surface tension in the simplest case \cite{kotaidis05apl,siems11}. However, it should be cautioned that the RPE is strictly valid only for a non-condensable gas in a liquid without additional internal (and temporally varying) heat source as represented by the hot nanoparticle. 
According to hydrodynamics in the RPE framework, the maximum bubble radius R$_{max}$ and the bubble energy E$_{bub}$ scale with \cite{linz25}:

\begin{eqnarray} \label{bubscale}
R_{max} &=& T/1.83 \cdot \sqrt(\frac{p_{stat}-p_v}{\rho _w})\\
E_{bub} & = & \frac{4 \pi}{3} R_{max}^3 (p_{stat}-p_v + 2 \sigma/R_{max})
\end{eqnarray}

with p$_{stat}$, p$_v$, $\rho_w$ and $\sigma$ representing ambient pressure, equilibrium vapor pressure at given temperature, water density and water surface tension, respectively, with T representing the lifetime of the bubble. 

Bubble formation has been observed in several studies, showing that with picosecond laser pulses bubble nucleation starts with applying fluences of 10 - 60 \Jm, depending on particle size, a range which overlaps with F$_0$ \cite{kotaidis05apl,kotaidis06jcp,siems11,metwally15,lombard17}. Figure \ref{bubble} shows the derived density change and temperature change in liquid water from fitting the liquid scattering differences. The density change in water compensates for growing vapor bubbles around the heated nanoparticles, while the total photo-excited water volume does not yet have time to expand because of the limited speed of sound. Thus, the measured density change of liquid water in the probed region is proportional to the total bubble volume \cite{kotaidis05apl}. In fig.  \ref{bubble} a) colloids containing 17 and 52 nm particles are compared. It is striking that the lifetime of the bubble changes from 2 ns for 17 nm to 5 ns for 52 ns, but the total density change is on a similar scale (note the lower fluence for the 17 nm particles). This shows that as the bubbles around 52 nm particles grow larger and thus live longer, the smaller particles of 17 nm are relatively higher in concentration, given that the atomic concentration is kept constant.  The bubble collapses at 5 ns for the 52 nm particles with a second peak showing up at 10 ns. RPE predicts bubble collapse followed by possible bubble revivals, which can easily be observed in single-bubble dynamics \cite{matsumoto13}, but is smoothed in a particle ensemble \cite{siems11,katayama14}. The simulations of Lombard et al. \cite{lombard16} have provided a more realistic picture of heat transport and bubble formation, which  explicitly includes excess stored energy in the nanoparticles. A consequence is a transition from heat conduction to ballistic heat flux in the vapor phase at nanometer distances. Additionally, several reviving bubbles may be fed by the still hot nanoparticles until a final, long-lived bubble appears. This would lead to an apparently delayed bubble formation time with the bubble size maximum being shifted to some nanoseconds despite the individual bubble lifetime of less than 1 ns. This behavior is observed in fig. \ref{bubble} b) where the bubble volumes of several particle sizes including 7 nm platinum particles are shown as a function of delay. One can observe that the growth time for the largest particles (57 nm) shifts up to 1 ns before a decay is observed up to 10 ns delay. Again, at a fixed fluence, the total bubble volume of all samples shows a similar amplitude, which indicates the scaling of bubble size with nanoparticle diameter. This is no surprise, as the total absorbance of the colloid is similar in all cases as noted for the threshold fluence F$_0$. Then, the total bubble volume reflects the total deposited heat as given in eq. \ref{eqheat}.

The temperature of the bulk water phase as extracted from the fit of the liquid scattering is also shown in fig. \ref{bubble} a). Basically, after release of the heat from the particles to the water phase after hundreds of nanoseconds, the temperature rises by a fraction of a Kelvin. A transient maximum in temperature at 2 ns for the 52 nm particles is explained by the compression of the water volume caused by bubble formation and consequently the temperature rise through adiabatic change of the state. The temperature rise with the 17 nm particles is smaller because of the lower fluence.

The bubble signal in fig. \ref{bubble} b is strongly reduced after 5 ns, but does not decay to zero within the first 20 nanoseconds. This consistent remainder is indicative of either further bubble revivals around the hot nanoparticles or creation of permanent gasses that dissolve on the nanosecond scale. An independent indication for that assumption is that the peak intensity of the (111) reflection (fig. \ref{bubble} c) vanishes within the time resolution of the experiment and only recovers slowly beginning at 5 ns to gain the final intensity at 30 ns. This coincidence of bubble vanishing and reappearance of crystallinity clearly shows the insulating character of the vapor bubbles around the particles, which slows down heat transport. With getting in contact to the liquid again the particles cool down rapidly and crystallize.  Note that the intensity does not recover fully after excitation at the fluence shown of 800 \Jm. This shows partial lattice destruction even after thermal relaxation, as will be discussed in the following section.

The bubble signal has been evaluated as a function of applied fluence in fig. \ref{bubble} d) at several delays, now displayed as a function of normalized  F/F$_0$. Bubble nucleation is found above the melting fluence with a clear bubble signal appearing at 800 ps at 1.4 F/F$_0$. Larger particles show bubble nucleation already below melting fluence (not shown) \cite{kotaidis06jcp,katayama14}. Interestingly, bubble signal at 5 ns is only detectable above 3 F/F$_0$, reflecting the prolongation of the bubble lifetime with increasing fluence according to eq. \ref{bubscale}. The bubble signal increases strongly with fluence to end in a stagnating amplitude at 6 F/F$_0$. 

\clearpage

\subsection{Particle fragmentation}

\begin{figure*}[ht] 
\caption{Results of the shape fits to the SAXS part of the difference scattering. a) Change of the total mass of the CTAB initial particles (lines) of 52, 47 and 17 nm diameter  and the total mass of the small fragmented particles (lines with symbols) as function of F/F$_0$. The grey area marks the prediction of change of the initial mass from the molecular dynamics simulations in \cite{plech24nano}. b) Diameter of the small fragmented particles in the CTAB samples as in a). c) Change of the total mass of the PLAL initial particles of 52, 47 and 17 nm diameter (lines) and the total mass of the small fragmented particles as function of F/F$_0$. The grey area marks the prediction of change of the initial mass from the molecular dynamics simulations in \cite{plech24nano}. d) Diameter of the small fragmented particles in the PLAL samples as in c).}
\label{frag}
\includegraphics[width=16cm,keepaspectratio]{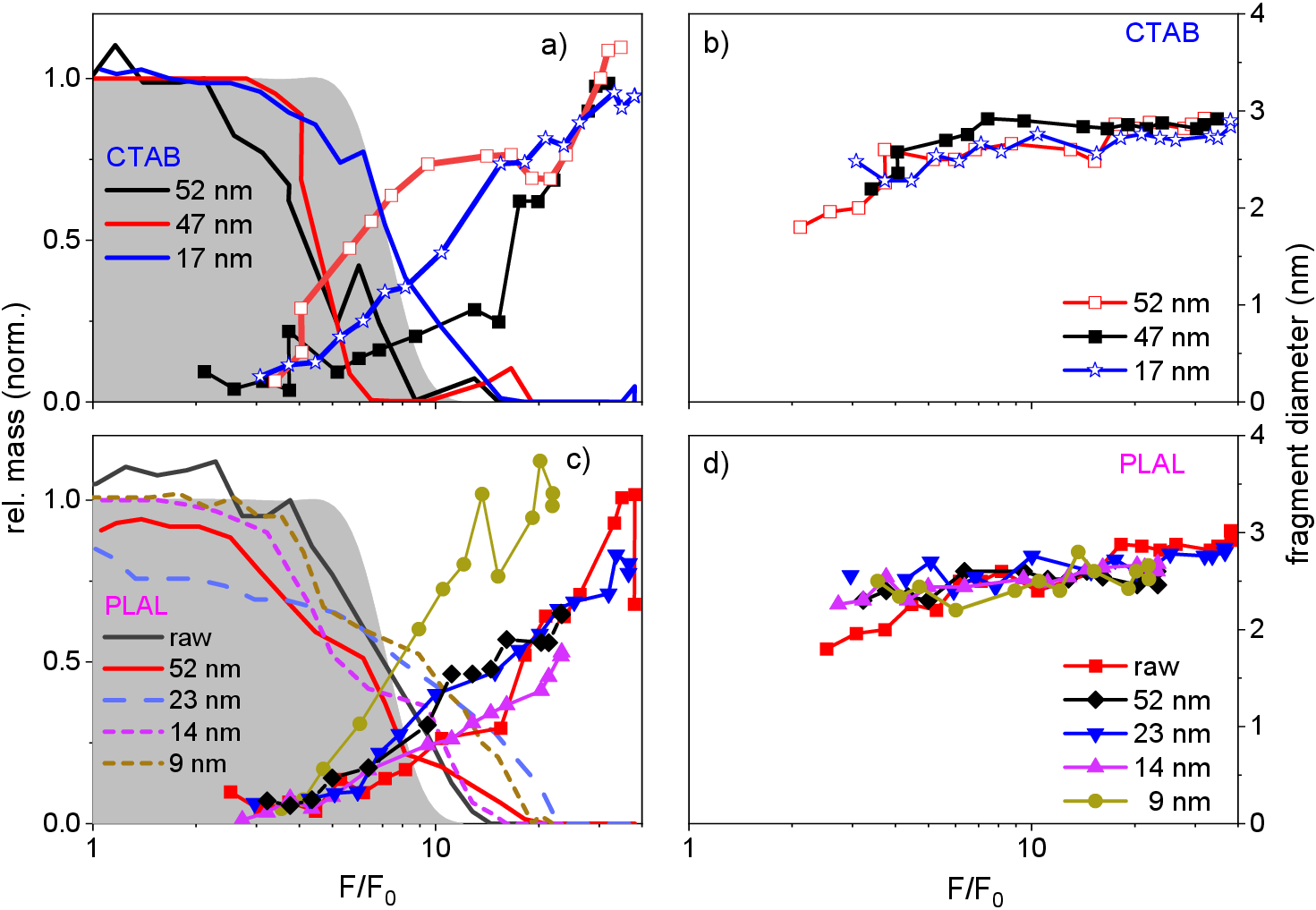}
\end{figure*}

With the lattice temperature being established and particle heating, melting, and bubble formation seen as a reversible reaction, one can analyze non-reversible structure changes as well. The SAXS part of the difference scattering shown in fig. \ref{sq} shows prominent changes t delay of 1 \mus, at which the sample has reached a stable configuration. No more changes are visible up to the largest measured delays of 20 \mus (not shown), where part of the excited sample volume starts to leave the probed position. Thus, a long-lasting/permanent change of particle shape is already found at 1.2 F/F$_0$. At this fluence, the difference scattering shows oscillations around zero, which are explained by a change of radius of gyration of CTAB particles. In fact, as the CTAB particles show a faceted shape after synthesis (mainly cuboids) the transient melting and recrystallization relaxes the shape into a sphere.  The observed changes can well be explained and simulated by this shape change. This signal is absent for PLAL particles, as these are typically crystallized from a molten droplet following ablation \cite{Shih2018}. At 4.8 F/F$_0$ in fig. \ref{sq} the oscillations are amplified and accompanied by a reduction of intensity at lowest q (0.01 \rezA). Additionally, at higher q (0.1 \rezA) a slight increase in scattering is detected. Already qualitatively, this is explained by a disappearance of the structure at a large length scale and the appearance of smaller sizes. This tendency is even amplified at 15 F/F$_0$. Quantitative fits, which are explained in detail in \cite{plech24nano} show the destruction of the initial particles after irradiation with the appearance of small clusters (< 4 nm, signal rise around 0.3 \rezA) as well as larger structures (signal rise around 0.1 \rezA). The larger structures are composed of larger  fragmented particles (> 4 nm) but also induced by positional correlation of the small particles. An indication for correlation is that the degree of stabilization (strong with CTAB, weak for PLAL particles) shows a different amplitude at this length scale. The total mass of the initial particles and the small fragments is displayed in fig. \ref{frag} a and c) for a range of different particle sizes. The first thing  to note is that the initial particles do not disappear below 4-5 F/F$_0$ and the reduction in mass is small above that threshold. . Only at 9-12 F/F$_0$ a strong reduction in the population of the initial particles is found. This is in agreement with the postulation of a thermal mechanism by which the particles heated slightly above the melting fluence would relax and recrystallize without changing size, but once the spinodal temperature is reached, spontaneous decomposition takes place via phase explosion. This scenario is predicted by the large-scale simulations carried out in \cite{huang21,huang22,plech24nano} with the change of initial particle mass resulting from the simulations being displayed as a shaded region in fig. \ref{frag}.

Strong phase explosion/fragmentation with the initial particle completely transformed into nanosized clusters is described at 12 F/F$_0$. Below 9 F$_0$ the simulations still show a limited size reduction, which is explained by evaporation from the surface, which has to be limited because of the short time span during which the particles stay hot. Note that the boiling temperature in the simulations is projected at 3.5 F/F$_0$ and full evaporation at 15 F/F$_0$. Indeed, the small fragments start to appear in fig. \ref{frag} a and c)  around 3 F/F$_0$, with significant mass relative to the initial particles only at 7-9 F/F$_0$. This general behavior is found throughout all investigated particle sizes. In particular, the PLAL particles (subfigure c) show a well comparable threshold fluence for fragmentation after scaling with the melting fluence. This is a strong indication for the proposed thermal mechanism, where temperature is the central order parameter. CTAB particles show a larger variation of fragmentation threshold fluence, with a lower onset. Although the evidence is low at the moment, given the statistical variations, one may speculate about the relevance of strain in fragmentation, as stress confinement increases with particle size \cite{spellauge23,pokhrel26}. Even more stringent is the coincidence of fragment sizes from various sample batches as shown in fig. \ref{frag} b and c). The fragments at low fluence have diameters of about 2 nm in each case, which is in agreement with the simulations. The simulations trace these smallest particles back to the evaporation from the surface with size growth as a function of delay. The diameters grow up to fluences of 10 F/F$_0$, where the strong phase explosion leads to spontaneous decomposition into fragments. In addition to the size-separated colloids for the PLAL particles we also processed the raw colloid directly from ablation with the same protocol. Fig. \ref{frag} shows that, scaled by the effective melting fluence the fragmentation process proceeds in the same fashion  quantitatively leading to the small fragments as the size-separated colloid samples. 

\clearpage

\subsection{Construction of a reaction map}

With the knowledge of all observed structural relaxations of the gold colloid it is noted that these laser-induced reactions are defined by laser fluence and time delay after excitation. After scaling with the threshold fluence for melting, several of the observed reactions become universal in F/F$_0$ even for colloids of different origin with varying physical properties. The total bubble volume  is also relative insensitive to particle size in the investigated range < 80-100 nm.  Temporally, the thermal phenomena depend on particle size, such as the thermal cooling of nanoparticles or the bubble typically scaling linearly. Thus, a generic reaction map can be composed in analogy to a phase diagram that maps the main pathways. 

\begin{figure*}[ht] 
\caption{Reaction map of the amplitude of lattice expansion (yellow) and (111) Bragg intensity (red) of 17 nm CTAB particles encoded by the diameter of the data points. The solid lines are guide to the eye to separate different regions as explained in the text. Note that above 10 F/F$_0$ the data points are sparse and cover only fluence-dependent data at delays of 5 ns, 20 ns, 100 ns and 1 \mus.}
\label{meltmap}
\includegraphics[width=16cm,keepaspectratio]{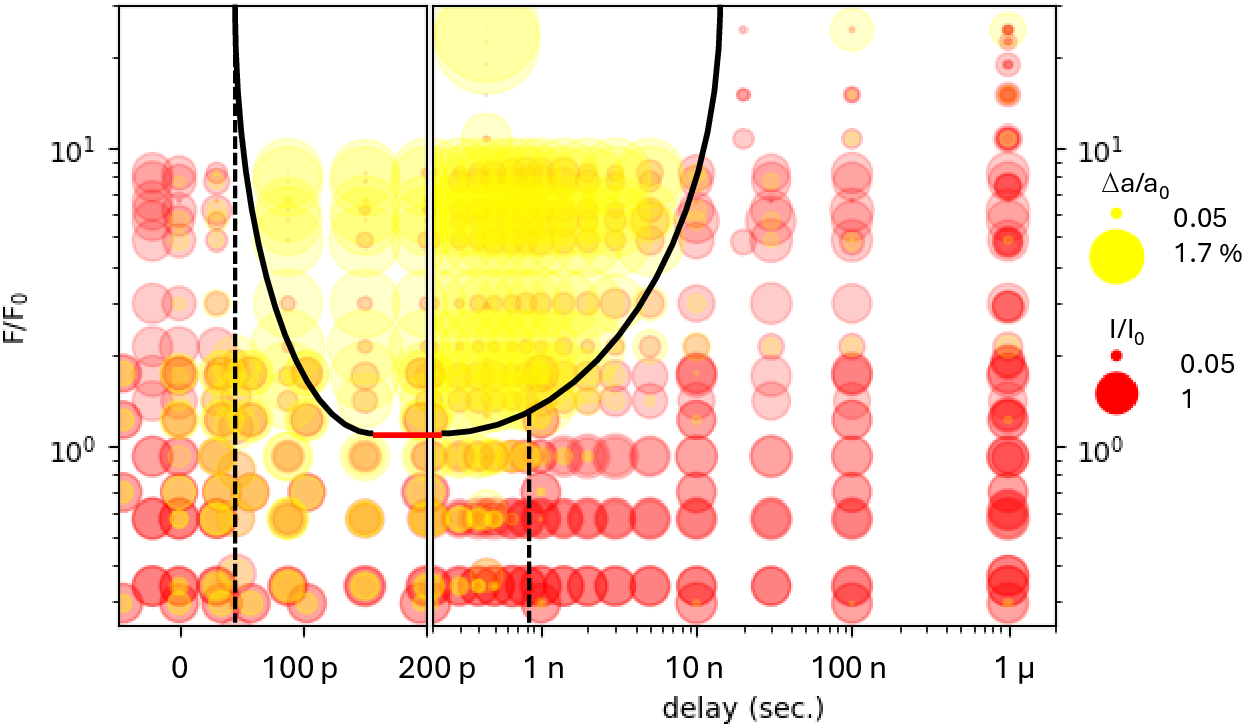}
\end{figure*}

Fig. \ref{meltmap} shows the 2D space of delay and reduced fluence F/F$_0$ for the intensity of the (111) powder peak of 17 nm CTAB particles and peak shift as indicators for melting and lattice temperature increase, respectively. Each data circle is one measurement, where the radius of the circles scales with the relative (111) intensity I/I$_0$ (red symbols) from 0 to 1 related to the intensity I$_0$ at negative delay, the lattice expansion $\Delta$a/a$_0$ (yellow symbols).   The particle heating sets in with the time resolution of the experiment and lasts for < 1 ns below the melting fluence. At the same time, the peak intensity is only reduced by a percentage as given by the Debye-Waller effect (see fig. \ref{expans} d). Above F$_0$ the peak intensity is strongly reduced  from 200 ps to 1 ns just above the melting fluence to extend to 40 ps to 10 ns at the highest fluence. The delayed melting has been discussed to originate from small overheating and slow nucleation of the molten phase. The area characterized by the molten state is encircled by the black line as a guide to the eye. The dashed lines show the onset of heating with the time resolution and the cooling below the melting fluence.   

\begin{figure*}[ht] 
\caption{Reaction map of the amplitude $\Delta \rho$ of bubble formation of 17 nm CTAB particles encoded by the size of the data points. The total bubble volume is inferred from the change of liquid scattering because of the compression at bubble formation. The solid lines are guide to the eye to separate different regions as explained in the text. Note that above 10 F/F$_0$ the data points are sparse as in fig. \ref{meltmap}.}
\label{bubmap}
\includegraphics[width=16cm,keepaspectratio]{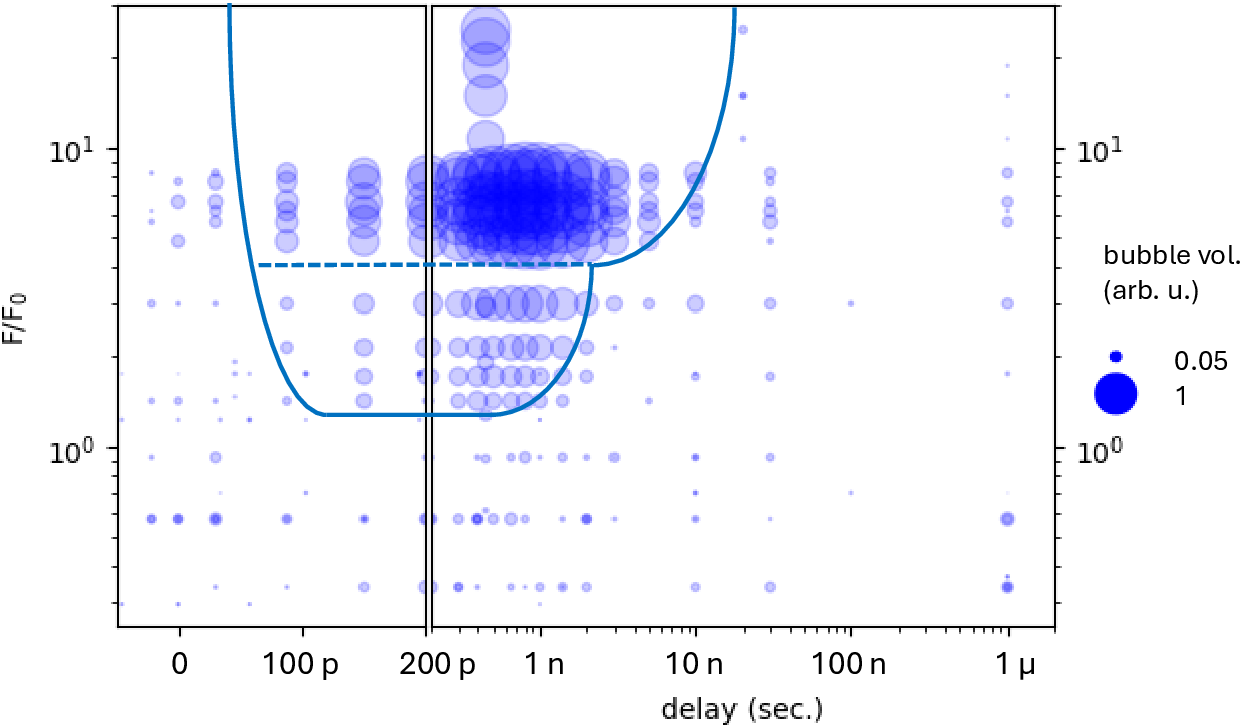}
\end{figure*}

Bubble formation is displayed in fig. \ref{bubmap}. As discussed above, the fluence threshold for bubble formation is slightly seen about F$_0$ with bubble sizes increasing with fluence. A bubble lifetime of 2.3 ns is found at moderate fluence. However, it is interesting to note that above 4 F/F$_0$ the bubble dynamics changes. The bubbles are seen to nucleate earlier and last for a longer delay of up to 10 ns. Therefore, we have tentatively divided the region of bubble appearance into a low-fluence and a high-fluence region. One origin of this change in behavior may be the transition between single-bubble cavitation and multiple-bubble revivals once the heat in the nanoparticles cannot be efficiently dissipated in few bubble oscillations \cite{lombard16}. A second explanation will be discussed below in relation to fragmentation.

\clearpage

\begin{figure*}[ht] 
\caption{Reaction map of the mass of fragment particles < 4 nm of 17 nm CTAB particles encoded by the size of the data points.  The lines separating different regions of heating, melting and bubble formation as shown in figs. \ref{meltmap}  and \ref{bubmap} are added. Note that above 10 F/F$_0$ the data points are sparse  as in fig. \ref{meltmap}.}
\label{fragmap}
\includegraphics[width=16cm,keepaspectratio]{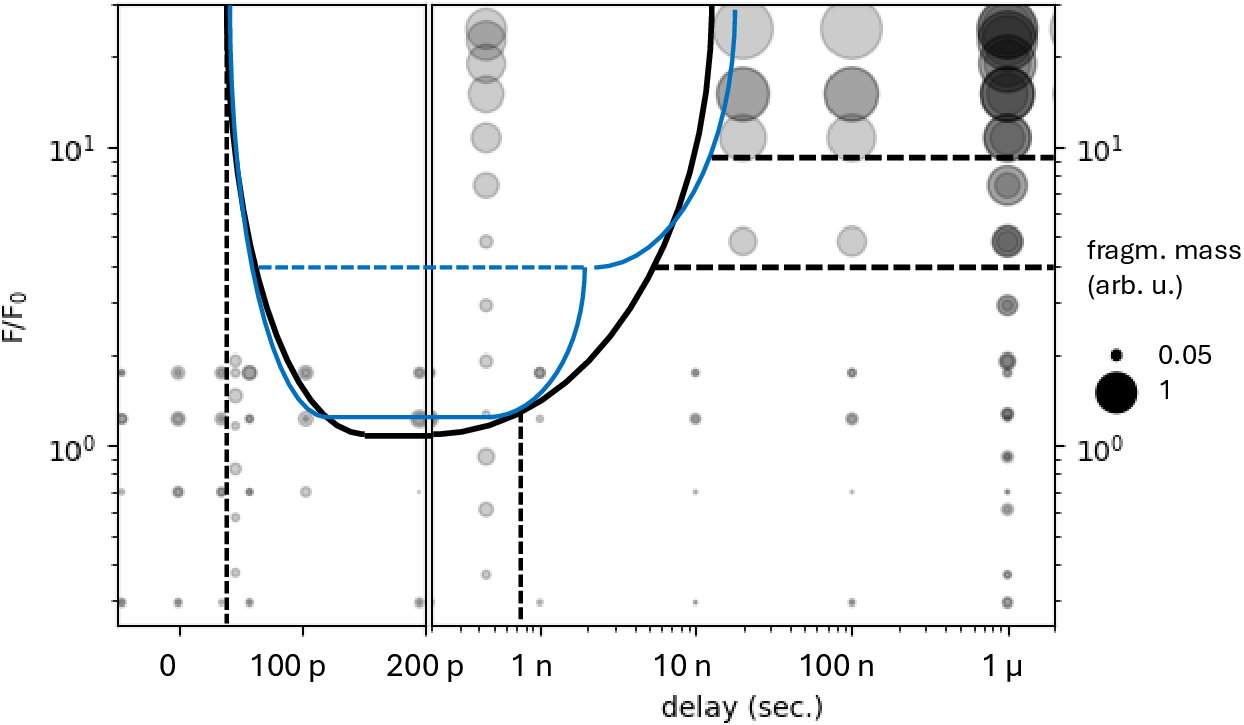}
\end{figure*}

Finally, the total mass of the fragmentation products is shown in fig. \ref{fragmap}  together with the guides to the eye from figs. \ref{meltmap} and \ref{bubmap}. The data show that the significant appearance of small particles is restricted to < 4 F/F$_0$, where surface evaporation and weak fragmentation enter, while the core of the initial particles still survives \cite{huang21,plech24nano}. Accordingly, the powder scattering seen in fig. \ref{meltmap} is slightly reduced, but vanishes only above 10 F/F$_0$, where a strong phase explosion disintegrates the entire particles to form small fragments. Already during the bubble phase, small particles can be detected, while resolution is limited because of the overlap of scattering signals. Earlier results \cite{ziefuss20} and manual inspection show that small particles are detectable at delays < 1 ns. At this delay, the particles surpass the detection limit of 1.4 nm of the present experiment. While the simulations show the genesis of small clusters within 100 ps these are small and thus not detected with the given signal-to-noise limit of the present experiment. Note that the scattering intensity in SAXS increases with volume as V$^2$. Only at 1 ns the growth of the particles surpasses sizes of 1.4 nm, which is the lowest detected size.  At later delays, the particles grow in size due to particle ripening and addition of monomers to the particles (fig. \ref{frag} c and d), which shows an apparent increase in particle mass as a function of delay. 

\subsection{Interplay of photoreactions}

Comparison of reaction maps of the different structural pathways in the excited colloid allows for correlating these pathways and inspection for interplay. Most prominently, the line of particle cooling and recrystallization above the fluence threshold for bubble formation coincide with the bubble lifetime. This allows us to conclude that the bubble is an envelope to the nanoparticles and shields them efficiently from cooling. Therefore, a final bubble collapse leads to rapid quenching of the temperature and cooling of the particles. Within the limitations of the experiment (time resolution and asynchronous rate-constant-like bubble collapse) we do not find indications for supercooling, but rather some spurious signal of warm crystalline particles after bubble collapse, in particular at the highest fluence in the strong fragmentation regime. Here, one could also question whether some clusters are still wrapped by a blanket of permanent gasses, which are known to be produced during laser fragmentation \cite{fromme25}  and laser ablation \cite{Kalus2017a}.

A correlation between fragmentation and bubble formation is apparent. At 4 F/F$_0$ weak fragmentation leads to separation of small particles, cluster species, and atoms from the surface of the initial particles \cite{huang22}, which coincides with the change of the bubble formation regime to small, short-lived bubbles to larger and long-lived bubbles. Molecular dynamics simulations by Huang and Zhigilei \cite{huang22} have revealed intriguing dynamics during the fragmentation process. At low energy input to the nanoparticle-water system (corresponding to roughly 4 F/F$_0$) only a surface part of the particles evaporates or is spalled off to end in the vapor bubble or at the interface between vapor and liquid water. With increasing energy, the structure of the fragmented particle undergoes a transient  change in the mechanism from hollow liquid sponge formation to the full fragmentation. An interesting phenomenon causes the hot spalled parts of the particle to be injected into the water phase and eventually be repelled to fuse to a larger particle again. This situation shows that at this point the speed of the spallation products is increased, which allows for  rapid release of energy into water. This, in turn, could lead to a more violent bubble formation process  seen here. 

Additionally, the question of whether fragmentation products can overcome the barrier at the vapor liquid interface or be pinned inside the bubble bears quite some importance for the stabilization of small clusters as opposed to ripening by close proximity. In analogy, it has been shown recently that the bubble confinement of particles in PLAL (on a millimeter scale) also causes an increased collision probability and leads to a ripening of particles timed by the bubble collapse \cite{Reich2017a,reich22}

\section{Conclusion and outlook}

\begin{figure*}[ht] 
\caption{Generalized reaction map as compiled from the figures \ref{meltmap}, \ref{bubmap} and \ref{fragmap} showing the region of particle thermal reactions and predominant structural motifs detected by time-resolved x-ray scattering. Detectable particle fragmentation coincides with the bubble stadium, which is reflected by the striped area between 100 ps and 20 ns above the threshold fluence for weak fragmentation.}
\label{scheme}
\includegraphics[width=17cm,keepaspectratio]{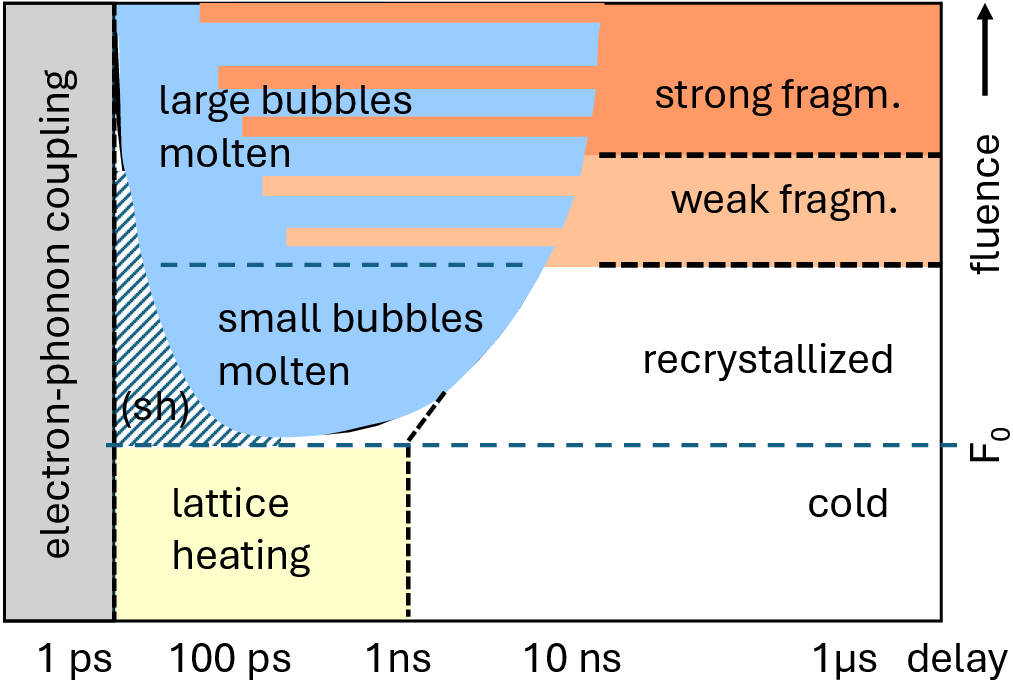}
\end{figure*}

Laser fragmentation in liquid by picosecond pulses has been studied in depth for the model system of gold nanoparticles in water. By employing pump-probe methods with x-rays as structural probe on a length scale from the atomic scale to a sub-micrometer length it is possible to register the photon-induced change of the particle lattice, particle shape as well as the reaction of the surrounding liquid phase to the rapid injection of heat on a sub-nanosecond scale. Particle heating, melting, and fragmentation can be discerned as a function of applied fluence and delay. Normalization of the fluence relative to an (extrapolated ) threshold fluence F$_0$ for particle melting allows to find the common energy scale for the reactions, in particular fragmentation into small particles < 4 nm. This is an important argument in classifying the fragmentation as a thermal process, as the mass evolution of initial particles and fragments as normalized fluence shows a very similar behavior regardless of particle size or colloid origin. In all cases, the particles survive the melting transition with subsequent crystallization unharmed. Fragmentation starts weakly only at 4 F/F$_0$  with full fragmentation achieved  above 9-12 F/F$_0$. This is explained by barrier-less decomposition by phase explosion, which is defined relative to particle temperature. Moreover, we have shown that because of this scaling effect and the weak variation of F$_0$ across a range of particle sizes, the fragmentation of a colloid as taken directly from PLAL with broad size distribution is  fragmented with an efficiency similar to size-selected colloids. This is encouraging for future high-throughput processing that directly uses the output from PLAL to further reducing sizes and narrow size distribution. A further consequence of the thermal-spinodal fragmentation is that particle size is not defined by the initial particles, but by the fragmentation mechanism itself, including decomposition into clusters and limited ripening thereafter. This evidenced by the matching fragment size distribution in fig. \ref{frag} b and d) for different initial samples including the growth from about 2 nm at fragmentation onset to less than 3 nm at high fluence. 

In this report, we have exemplarily shown reaction map data for one specific colloid sample. Maps of other colloid sizes and synthesis protocol show the same features of particle melting, recrystallization, bubble formation, and fragmentation. This has been detailed in the comparison of the individual measurements. Bubble formation in particular shows a shift in time scale with particle size as larger particles absorb more heat on an absolute scale, which in turn increases the bubble size and life time correspondingly. To first approximation the bubble size and its lifetime scale linearly with particle size superheating of the solid particles beyond the melting fluence, contrarily is a phenomenon that depends critically on particle morphology, such as size or crystallinity.

The comparability of reactions for different colloids leads to the notion that a generic reaction map can be constructed, which shows the dominant reaction as a function of delay and fluence. Such a scheme is presented in fig. \ref{scheme}.

\begin{funding}
The following sources of funding are acknowledged: 
German Science foundation DFG for grants 491072288 and by the Helmholtz Association program “From Matter to Materials and Life.” (A. P.). We acknowledge the European Synchrotron Radiation Facility for provision of beam time on ID09; 

\end{funding}

\end{document}